\documentclass[conference,10pt]{IEEEtran}
\usepackage{xcolor,soul,framed} 

\colorlet{shadecolor}{yellow}
\usepackage[
pdftex]{graphicx}
\graphicspath{{../pdf/}{../jpeg/}}

\DeclareGraphicsExtensions{.pdf,.jpeg,.png}

\usepackage[cmex10]{amsmath}
\usepackage{array}
\usepackage{mdwmath}
\usepackage{mdwtab}
\usepackage{eqparbox}
\usepackage{url}
\usepackage{amsmath}
\usepackage{amssymb}
\usepackage{graphicx}
\usepackage{tabularx}
\usepackage{subcaption}
\usepackage{placeins}

\begin{document}
\bstctlcite{IEEEexample:BSTcontrol}
    \title{Quantum Enhanced Entropy Pool for Cryptographic Applications and Proofs}
  \author{
    \IEEEauthorblockN{
    Buniechukwu Njoku\IEEEauthorrefmark{1},
    Sonai Biswas\IEEEauthorrefmark{1},
    Milad Ghadimi\IEEEauthorrefmark{1},
    Mohammad Shojafar\IEEEauthorrefmark{3}, 
    Gabriele Gradoni\IEEEauthorrefmark{3}
    Riccardo Bassoli\IEEEauthorrefmark{1},\\
    Frank H. P. Fitzek\IEEEauthorrefmark{1}\IEEEauthorrefmark{2}
    }
    
    \IEEEauthorblockA{
    \IEEEauthorrefmark{1}Deutsche Telekom Chair of Communication Networks, Technische Universität Dresden, Dresden, Germany\\
    \IEEEauthorrefmark{2}Centre for Tactile Internet with Human-in-the-Loop (CeTI), Cluster of Excellence, Dresden, Germany\\
    \IEEEauthorrefmark{3}5G/6GIC, Institute for Communication Systems (ICS), University of Surrey, Guildford, UK\\
    \{buniechukwu\_chidike.njoku, sonai.biswas, milad.ghadimi, riccardo.bassoli, frank.fitzek\}@tu-dresden.de\\ \{m.shojafar, g.gradoni\}@surrey.ac.uk\\
    }
}


\markboth{ IEEE Journal on Selected Areas in Communications, VOL.~60, NO.~12, JULY~2024
}{Roberg \MakeLowercase{\textit{et al.}}: Secure VRFs for Networked Systems}

\maketitle

\begin{abstract}

This paper investigates the integration of quantum randomness into Verifiable Random Functions (VRFs) using the Ed25519 elliptic curve to strengthen cryptographic security. By replacing traditional pseudorandom number generators with quantum entropy sources, we assess the impact on key security and performance metrics, including execution time, and resource usage. Our approach simulates a modified VRF setup where initialization keys are derived from a quantum random number generator source (QRNG). The results show that while QRNGs could enhance the unpredictability and verifiability of VRFs, their incorporation introduces challenges related to temporal and computational overhead. This study provides valuable insights into the trade-offs of leveraging quantum randomness in API-driven cryptographic systems and offers a potential path toward more secure and efficient protocol design. The QRNG-based system shows increased (key generation times from 50 to 400+ µs, verification times from 500 to 3500 µs) and higher CPU usage (17\% to 30\%) compared to the more consistent performance of a Go-based VRF (key generation times below 200 µs, verification times under 2000 µs, CPU usage below 10\%), highlighting trade-offs in computational efficiency and resource demands.

\end{abstract}


%
\IEEEpeerreviewmaketitle


\section{Introduction}

In the rapidly advancing field of cryptography, the integration of quantum resources represents a profound opportunity for achieving unprecedented levels of security and robustness. At the core of this advancement lies true randomness, a fundamental requirement for secure key generation in cryptographic primitives such as Verifiable Random Functions (VRFs) and Elliptic Curve Cryptography (ECC). The ability to generate highly unpredictable, high-entropy keys and nonces from quantum sources is essential for safeguarding cryptographic systems against a broad spectrum of attack vectors, including side-channel attacks that exploit indirect information leaks such as power consumption, timing variations, and electromagnetic emissions\cite{simion2020entropy}.

Quantum Random Number Generators (QRNGs) offer a transformative approach to enhancing cryptographic security by leveraging the intrinsic unpredictability of quantum mechanical processes. Unlike classical pseudorandom number generators (PRNGs), which are inherently deterministic and vulnerable to various predictive attacks, QRNGs produce randomness that is fundamentally non-deterministic and highly resistant to replication or prediction. \cite{chowdhury2021physical}. This quality makes QRNGs a powerful complementary enhancement tool for generating cryptographic keys to support the underlying cryptographic algorithms, and additional countermeasures required to mitigate threats like side-channel attacks and other sophisticated attack methods.

\noindent\textbf{Main Contributions:} This paper comprehensively explores integrating QRNGs within VRF systems, particularly those based on the Ed25519 elliptic curve, to rigorously assess the trade-offs between enhanced security and the potential impacts on performance. By incorporating quantum-generated randomness into the key generation process, we aim to demonstrate the potential for substantial improvements in cryptographic strength while critically evaluating the practical challenges associated with latency, execution time, and resource utilization in API-driven cryptographic environments. Through this investigation, we provide a detailed analysis of the optimization strategies necessary for deploying cryptographic systems that harness quantum randomness, thereby paving the way for developing more secure, efficient, and resilient application-layer protocols.

\noindent \textbf{Organization: }The rest of the paper is structured as follows. Section~\ref{sec:2} discusses related work on VRF and ECC encryption methods focusing on advancements in quantum-resistant cryptography techniques. Section~\ref{sec:3} presents the proposed system. Section~\ref{sec:4} provides a comprehensive performance evaluation of the QRNG model used for the VRF system. Finally, Section~\ref{sec:5} concludes the paper and suggests potential directions for future research.

\section{Background}\label{sec:2}

Verifiable Random Functions (VRFs), an extension of the Goldreich-Goldwasser-Micali random functions, are critical cryptographic primitives that produce publicly verifiable yet unique outputs\cite{814584}. Traditional VRF implementations, which rely on RSA and elliptic curve cryptography (ECC), base their security on the hardness of the elliptic curve discrete logarithm problem (ECDLP). Among these, the Edwards-curve Digital Signature Algorithm (EdDSA) \cite{nikolaenko2021non}, particularly its variant Ed25519 based on Curve25519, has gained prominence due to its high efficiency, performance, and robust security properties. Congruently, there has been increasing interest in Post-Quantum Cryptography (PQC), which includes cryptographic algorithms resistant to quantum attacks. Lattice-based cryptography \cite{nejatollahi2019post}, hash-based signatures \cite{li2022hash}, and multivariate polynomial cryptography \cite{10.1145/3571071} are some of the leading candidates for PQC. These algorithms are being developed as part of the NIST Post-Quantum Cryptography Standardization project \cite{alagic2022status}, which aims to identify quantum-resistant cryptographic methods, with QRNGs remaining highly relevant in this context. 
Verifiable Random Functions (VRFs) have diverse applications across various domains, particularly in communication systems and internet security. One prominent use case is in the Domain Name System Security Extensions (DNSSEC) protocol, where VRFs provide cryptographic proofs for non-existence responses in DNSSEC, preventing attackers from enumerating valid domain names. For instance, the NSEC5 protocol\cite{goldberg2016nsec5, papadopoulos2017making} leverages VRFs to ensure the integrity and privacy of domain name responses. Additionally, VRFs play a crucial role in key transparency systems, such as Google Key Transparency\cite{google_key_transparency_vrf}, CONIKS\cite{coniks_vrf_go}, and Yahoo!'s Coname\cite{ishiguro_vrf}, where they enable users to verify the authenticity of public keys while preserving privacy. In the blockchain ecosystem, VRFs are vital for enhancing the fairness and security of consensus protocols by generating unpredictable and tamper-resistant random values. For example, Algorand's Pure Proof-of-Stake (PPoS) consensus mechanism\cite{esgin2021practical} employs VRFs to privately and randomly select block proposers and validation committees, thereby preventing manipulation and ensuring scalability and security. Similarly, Cardano’s Ouroboros Proof-of-Stake protocol\cite{badertscher2022uc} uses VRFs to securely determine which nodes are eligible to produce the next block. Beyond consensus mechanisms, VRFs also find applications in other areas of blockchain technology. Chainlink\cite{breidenbach2021chainlink}, for example, utilizes VRFs to provide tamper-proof randomness for smart contracts.
The security of elliptic curve cryptography has been extensively studied, particularly its resilience against side-channel attacks. For example, in \cite{10.1007/978-3-642-10366-7_39}, vector quantization and Hidden Markov Model (HMM) analysis of cache access times have shown that template attacks \cite{chari2003template, medwed2008template} can model the internal state of a system and ultimately expose private keys. In contrast, quantum randomness, derived from inherently unpredictable quantum processes, provides a stronger foundation for secure key generation\cite{ma2016quantum}. Unlike classical pseudorandom number generators (PRNGs), which rely on deterministic algorithms and require careful seeding to ensure unpredictability, Quantum Random Number Generators (QRNGs) produce true randomness that is fundamentally non-deterministic. This provides a stronger guarantee of entropy, as the randomness originates from physical processes governed by quantum mechanics. Among the established methods, quantum dots offer a robust means of generating entangled photon pairs, producing high-quality random numbers essential for cryptographic applications \cite{jons2017bright, huber2018strain,wang2019demand}. 
The security of a VRF relies on several core properties. One such property is \textit{Domain-Range Correctness} \cite{814584, abdalla2014verifiable}, which ensures that for all inputs \( x \in \mathcal{D} \), the output \( F_{\text{sk}}(x) \in \mathcal{R} \) holds with overwhelming probability over the choices of the key pairs \( (pk, sk) \). Another key property is \textit{Unique Provability} \cite{814584, abdalla2014verifiable}, which requires that for every public key \( pk \), input \( x \), and values \( v_1, v_2 \) with corresponding proofs \( \pi_1, \pi_2 \), such that \( v_1 \neq v_2 \), the probability that \( \text{Verify}(pk, x, v_i, \pi_i) = \text{YES} \) for both \( i \in \{1, 2\} \) is negligible.

\section{Methodology}\label{sec:3}

This paper presents a system that integrates Quantum Random Number Generators (QRNGs) into the implementation of Verifiable Random Functions (VRFs) using the Ed25519 elliptic curve, with a primary focus on evaluating performance metrics. By replacing traditional pseudorandom number generators (PRNGs) with quantum-generated entropy sources, we investigate the impact of this integration on key performance factors such as execution time and resource utilization.
This study evaluates the performance implications of using QRNGs for cryptographic key generation in Elliptic Curve Cryptography (ECC) and VRF systems. We conduct a series of simulations to analyze how quantum randomness affects these performance metrics, particularly in the context of API designs tailored for application-layer protocols. While classical cryptographic PRNGs can generate secure random numbers under well-established assumptions, QRNGs offer a source of true randomness, potentially providing stronger entropy guarantees in adversarial settings. However, this may introduce practical trade-offs, which we evaluate through performance analysis.
Our approach involves using QRNGs to generate private keys for EdDSA and VRF implementations, with a focus on assessing the practical performance implications of integrating quantum randomness. Specifically, we examine the impact of a QRNG on cryptographic operations in comparison to a classical PRNG. This section details our methodology for incorporating QRNGs and provides an in-depth analysis of their impact on system performance, particularly concerning API design considerations for application-layer protocols.

\subsection{Ed25519 and its Relevance to VRFs}

Ed25519 leverages the elliptic curve Curve25519, which operates over a prime field and provides approximately 128-bit security for digital signatures under standard cryptographic assumptions, which is suitable for modern cryptographic applications. Designed by Daniel J. Bernstein, Curve25519 offers high security, fast computation, and resistance to side-channel attacks, making it ideal for environments where both performance and security are critical.

The Ed25519 signature scheme involves the following steps\cite{nikolaenko2021non}:

1. \textit{Key Generation}: A private key seed is randomly selected and hashed to derive the scalar \( s \in \mathbb{Z}_q \), where \( q \) is a prime number that defines the order of the base point \( B \) on the elliptic curve. The public key \( pk \) is computed as:
   \begin{equation}
    pk = sB
   \end{equation}
   where \( B \in G \) is the designated base point of the group \( G \) in which the discrete logarithm problem is hard.

   In the modified VRF system, the selection of the private key seed \( s \) is done using a \textit{Quantum Random Number Generator (QRNG)}, replacing the traditional pseudorandom number generator (PRNG). Specifically, the seed \( s \) is generated by reading a stream of high-entropy quantum randomness from external binary files using a custom \texttt{BitReader}. 
   
2. \textit{Signing}: To sign a message \( m \), a scalar \( r \in \mathbb{Z}_q \) is deterministically derived from the private key and the message itself, ensuring that the signature is unique for each message. The signature \( \sigma \) consists of the pair \( (R, S) \), where:
   \begin{equation}
   R = rB, \quad S = r + \mathcal{H}(R, pk, m) s \pmod{q},
   \end{equation}
   and \( \mathcal{H} \) is a cryptographic hash function.

3. \textit{Verification}: To verify a signature \( \sigma = (R, S) \) on a message \( m \) with public key \( pk = s \cdot B \), the verifier checks the following equation:
   \begin{equation}
   S B = R + \mathcal{H}(R, pk, m) pk,
   \end{equation}
   This verification equation relies on the assumption that computing the discrete logarithm is infeasible, thereby ensuring the integrity and authenticity of the signed message.

The unpredictability and verifiability of a Verifiable Random Function (VRF) critically depend on the quality of the entropy source used to generate the seed \( s \). If the entropy source is low or biased, the function \( f(x) = F(s, x) \) becomes predictable, compromising security. Therefore, to maximize entropy, a \textit{Quantum Random Number Generator (QRNG)} is recommended, ensuring high randomness and entropy.

The Elliptic Curve Verifiable Random Function (ECVRF) follows a structured process\cite{10.1007/978-3-031-30872-7_4}:

1. \textit{Key Generation (Gen)}: A secret key \( x \in \mathbb{Z}_q \setminus \{0\} \) is selected. The corresponding public key is:
   \begin{equation}
   X = B^x \in G,
   \end{equation}
   where \( G \) is a cyclic group of prime order \( q \) with generator \( B \).

   In the QRNG-modified system, the secret key \( x \) is selected by generating the seed using the \texttt{BitReader}, which extracts high-entropy randomness from QRNG-derived binary files. This process replaces the standard PRNG method, ensuring that the scalar \( x \) is generated from an unassailable source of randomness.

2. \textit{Hash to Curve (HTC)}: Each VRF input \( \alpha \in X \) is mapped to a point \( P \in G \) using a hash function \( \text{HTC} \):
   \begin{equation}
   P := \text{HTC}(X, \alpha) \in G.
   \end{equation}

3. \textit{Computation of VRF Output}: The prover computes:
   \begin{equation}
   Z := P^x \in G.
   \end{equation}
   A proof \( \pi = (Z, c, s) \) is constructed, where:
   \begin{align}
   &\text{- A random value } r \in \mathbb{Z}_q \text{ is chosen.} \notag \\
   &\text{- Commitments are calculated:} \notag \\
   &R_B := B^r \in G, \quad R_P := P^r \in G. \notag
   \end{align}
   A challenge is computed:
   \begin{equation}
   c := \mathcal{H}(P, Z, R_B, R_P) \in H.
   \end{equation}
   The response is:
   \begin{equation}
   s := r + x \cdot c \in \mathbb{Z}_q.
   \end{equation}

   In the modified implementation, the random scalar \( r \) is generated using input from the \textit{QRNG-sourced randomness}. This replaces the conventional randomness generation process (typically via \texttt{rand.Reader}) with quantum entropy. This ensures the unpredictability of the scalar \( r \), which plays a critical role in constructing a secure proof.

4. \textit{Verification (Verify)}: Given public key \( X \in E \), input \( \alpha \in X \), and proof \( \pi = (Z, c, s) \), the verifier calculates:
   \begin{equation}
   P := \text{HTC}(X, \alpha) \in G.
   \end{equation}
   The verifier checks:
   \begin{equation}
   R_B := B^s X^{-c} \in G, \quad R_P := P^s Z^{-c} \in G.
   \end{equation}
   If:
   \begin{equation}
   c = \mathcal{H}(P, Z, R_B, R_P),
   \end{equation}
   the output is valid; otherwise, it outputs an invalid value.

\section{Results and Discussions}\label{sec:4}
This study aimed to highlight the performance trade-offs of the QRNG-based system compared to a VRF implemented in Go, providing insights for future API design. We tested 10 million proofs generated from the local random number generator native to Go(rand), against keys from ANU's QRNG(qrng). The VRF implemented is based on a fork of NKN's full node implementation \cite{nknorg_nkn}, and the random keys were sourced from ANU's QRNG\cite{haw2015maximization}. The simulations were conducted on a Ryzen 7 PRO 7840U processor featuring Radeon 780M Graphics. This CPU has 8 cores and 16 threads, with a base clock speed of 1.9 GHz and a maximum turbo speed of 5.1 GHz. It includes an L1 cache of 512 KiB, an L2 cache of 8 MiB, and an L3 cache of 16 MiB, with 32 GiB of DDR5 memory, arranged as four 8 GiB modules, operating at an effective clock speed of 2105 MHz.
The system monitored memory usage using the Go \texttt{runtime.MemStats} function, which allowed for dynamic tracking of memory allocation and deallocation during cryptographic tasks such as key generation, VRF evaluation, and proof generation. Furthermore, CPU utilization was monitored via the \texttt{gopsutil} library, which tracked real-time CPU usage, enabling the system to adjust and balance computational load dynamically. These metrics were logged every 100 operations, offering granular insights into system performance.

\subsection{Latency Estimation}

\begin{figure}[htbp]
    \centering
    \includegraphics[width=1\linewidth]{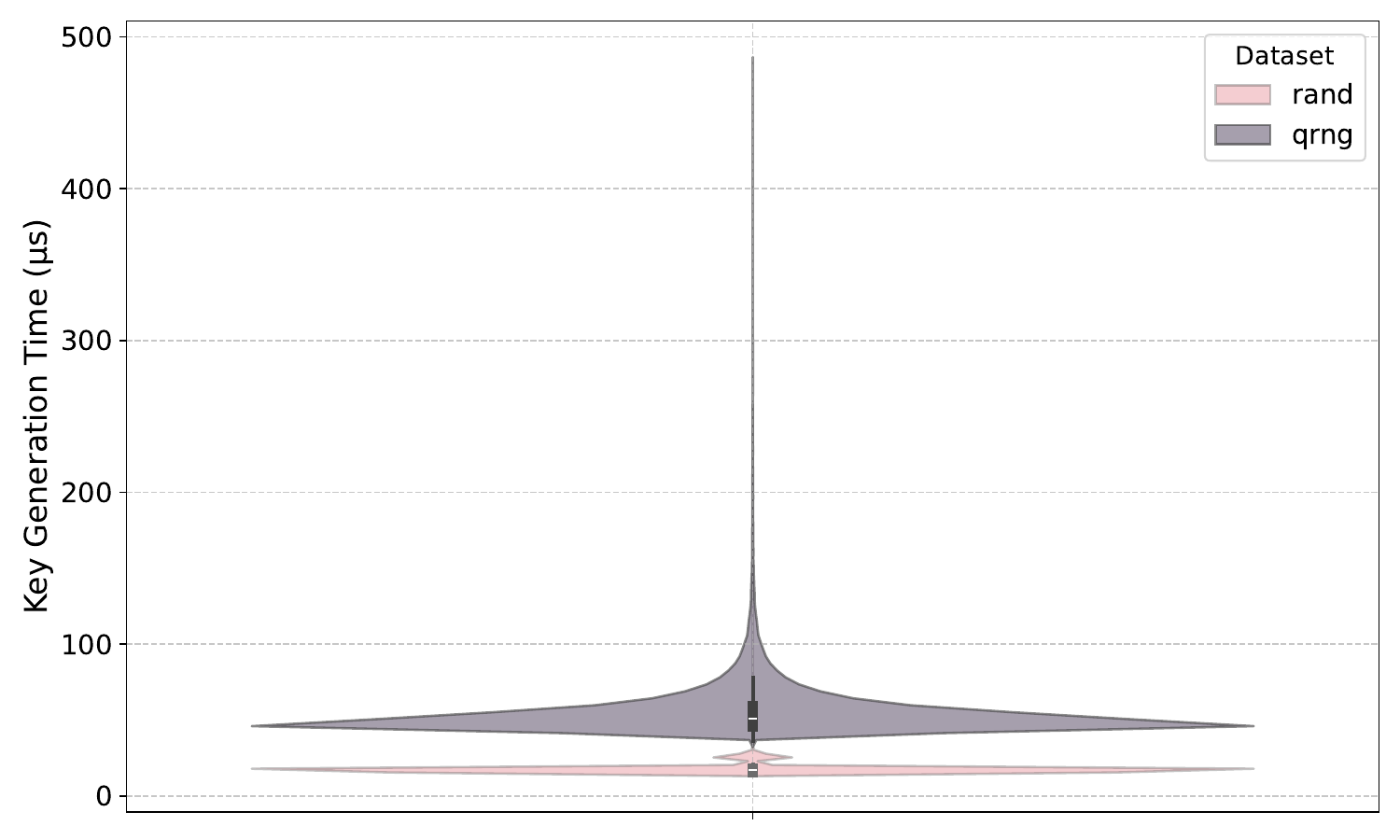}
    \caption{Key Generation Time}
    \label{fig:violin_keygen}
\end{figure}

The qrng system showed a broad range in key generation times, from 50 to 150 µs for smaller datasets, but with spikes exceeding 400 µs as the dataset size surpasses 8 million elements (e.g., 420 µs at 8 million elements), indicating potential inefficiencies at larger scales (Figure 1). In contrast, rand maintained more stable key generation times, consistently below 200 µs, even at 10 million elements, demonstrating more efficient scaling (Figure 1, right).
\begin{figure}[htbp]
    \centering
    \includegraphics[width=1\linewidth]{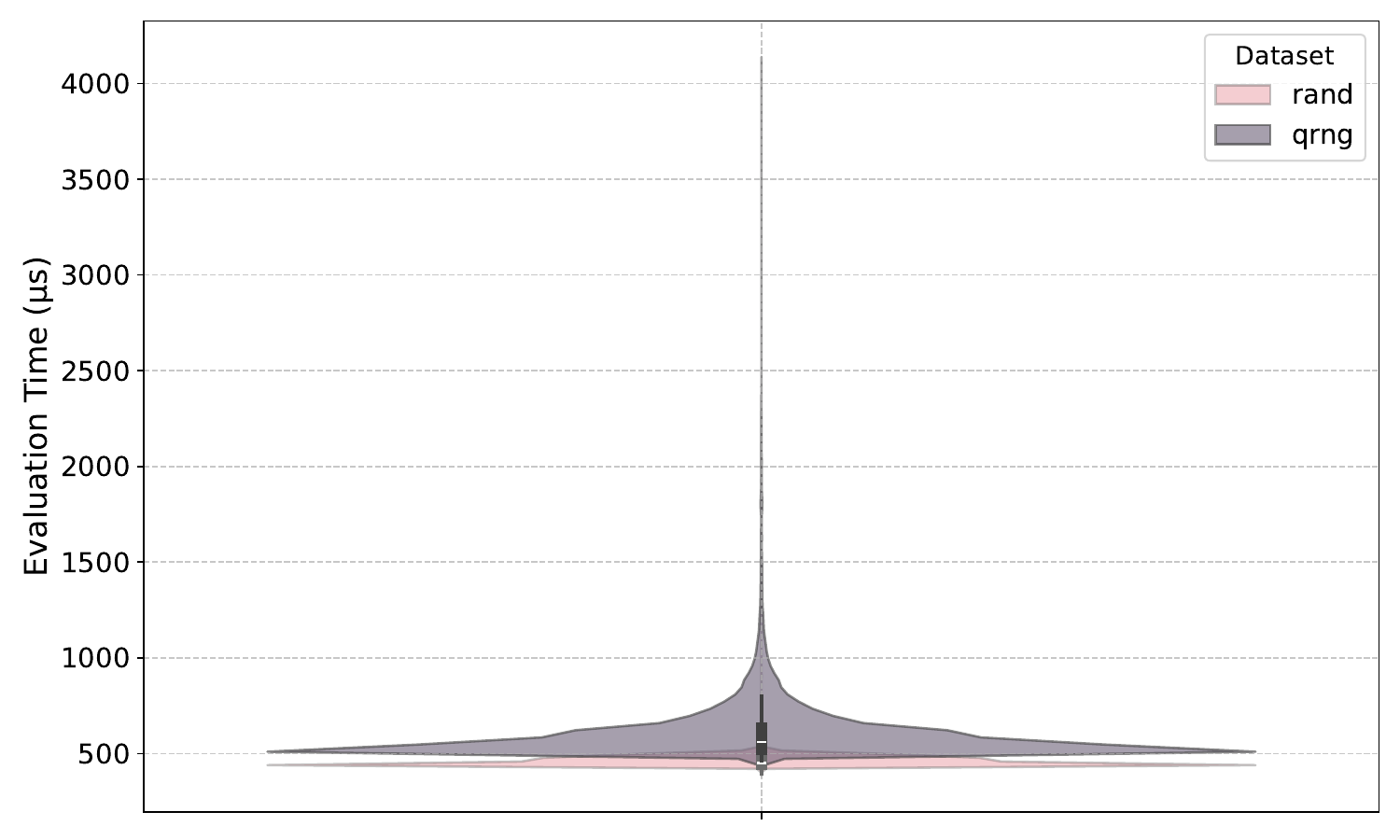}
    \caption{Evaluation Time}
    \label{fig:violin_eval}
\end{figure}

qrng evaluation times varied from 500 µs to over 3000 µs, with increased variance at larger dataset sizes (e.g., 3500 µs at 10 million elements), suggesting challenges in maintaining performance (Figure 2). The rand VRF showed a narrower range (500–2000 µs), with consistent times around 1000 µs even for large datasets, reflecting better scalability (Figure 2).

\begin{figure}[htbp]
    \centering
    \includegraphics[width=1\linewidth]{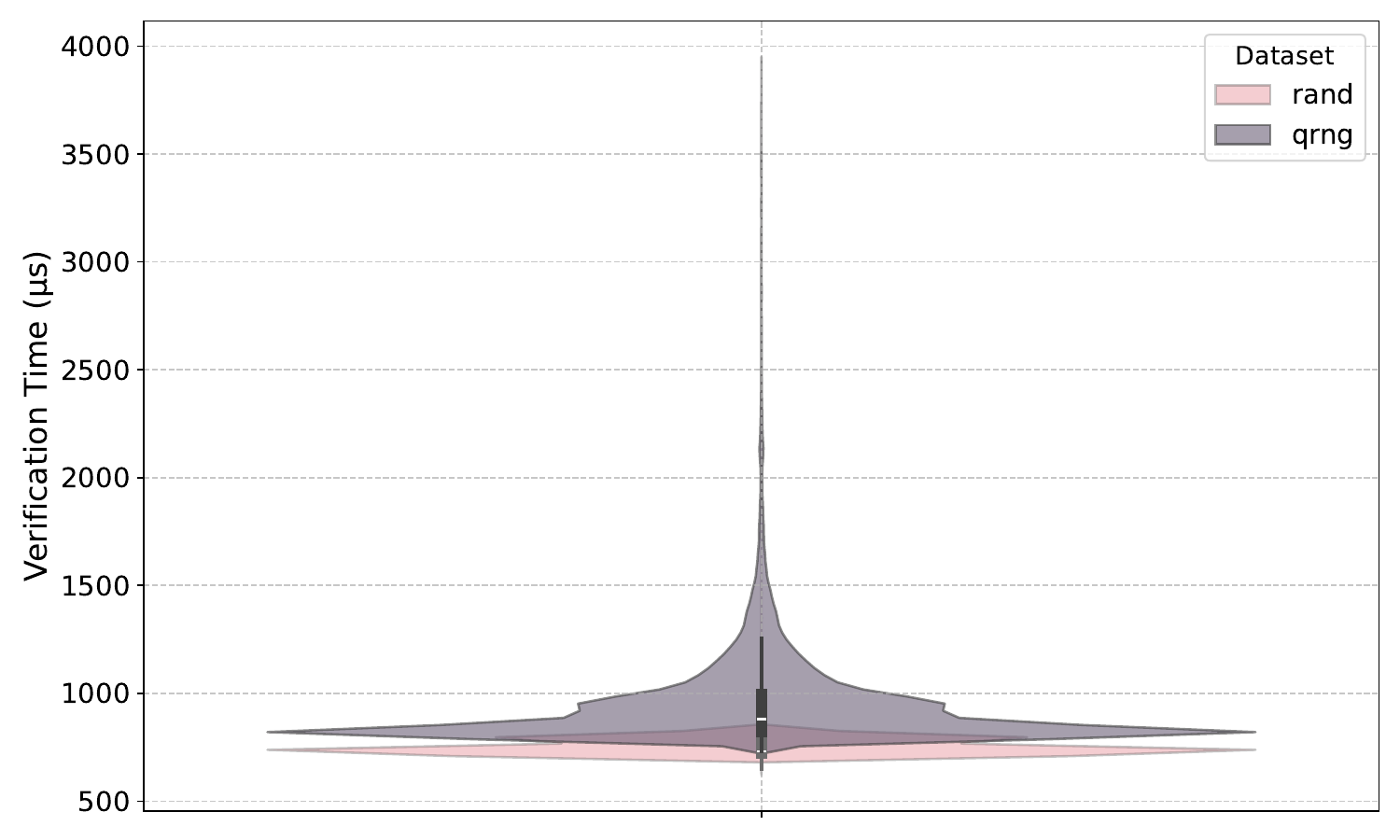}
    \caption{Verification Time}
    \label{fig:violin_verif}
\end{figure}

Verification times for qrng fluctuated between 500 µs and 3500 µs, with higher values observed as datasets grew (e.g., 3200 µs at 9 million elements), potentially affecting predictability (Figure 3). The rand system, however, maintained verification times mostly below 2000 µs, offering greater stability (Figure 3).

\begin{figure}[htbp]
    \centering
    \includegraphics[width=1\linewidth]{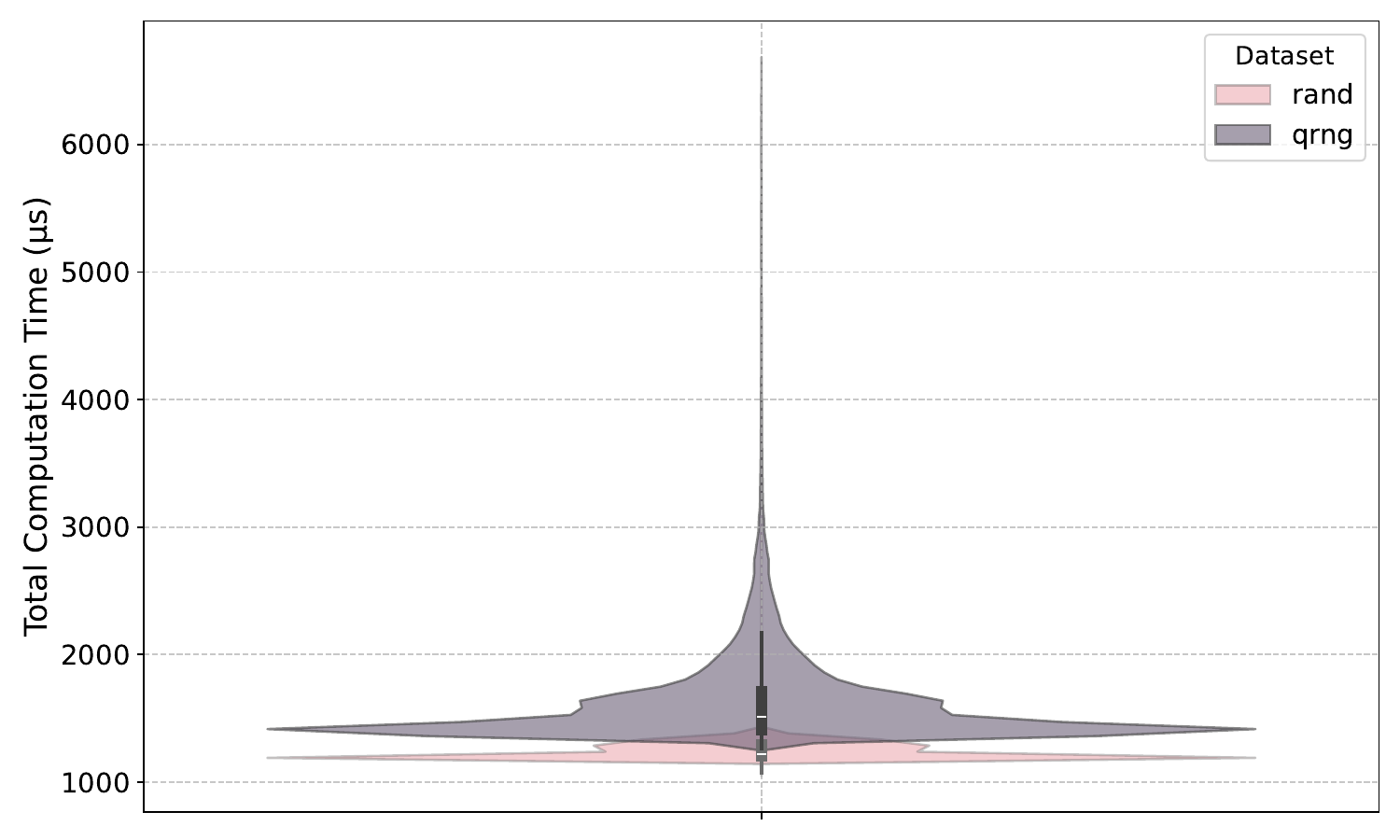}
    \caption{Total Computation Time}
    \label{fig:violin_tcomp}
\end{figure}

Total computation time for qrng ranged from 1000 µs to over 6500 µs, with notable spikes beyond 8 million elements (e.g., 6100 µs at 9 million elements), suggesting scalability limitations (Figure 4). The rand VRF maintained a more consistent range (1000–4000 µs), with fewer outliers, indicating better optimization for larger datasets (Figure 4).

\subsection{Computational Resource Estimation}

\begin{figure}[htbp]
    \centering
    \includegraphics[width=1\linewidth]{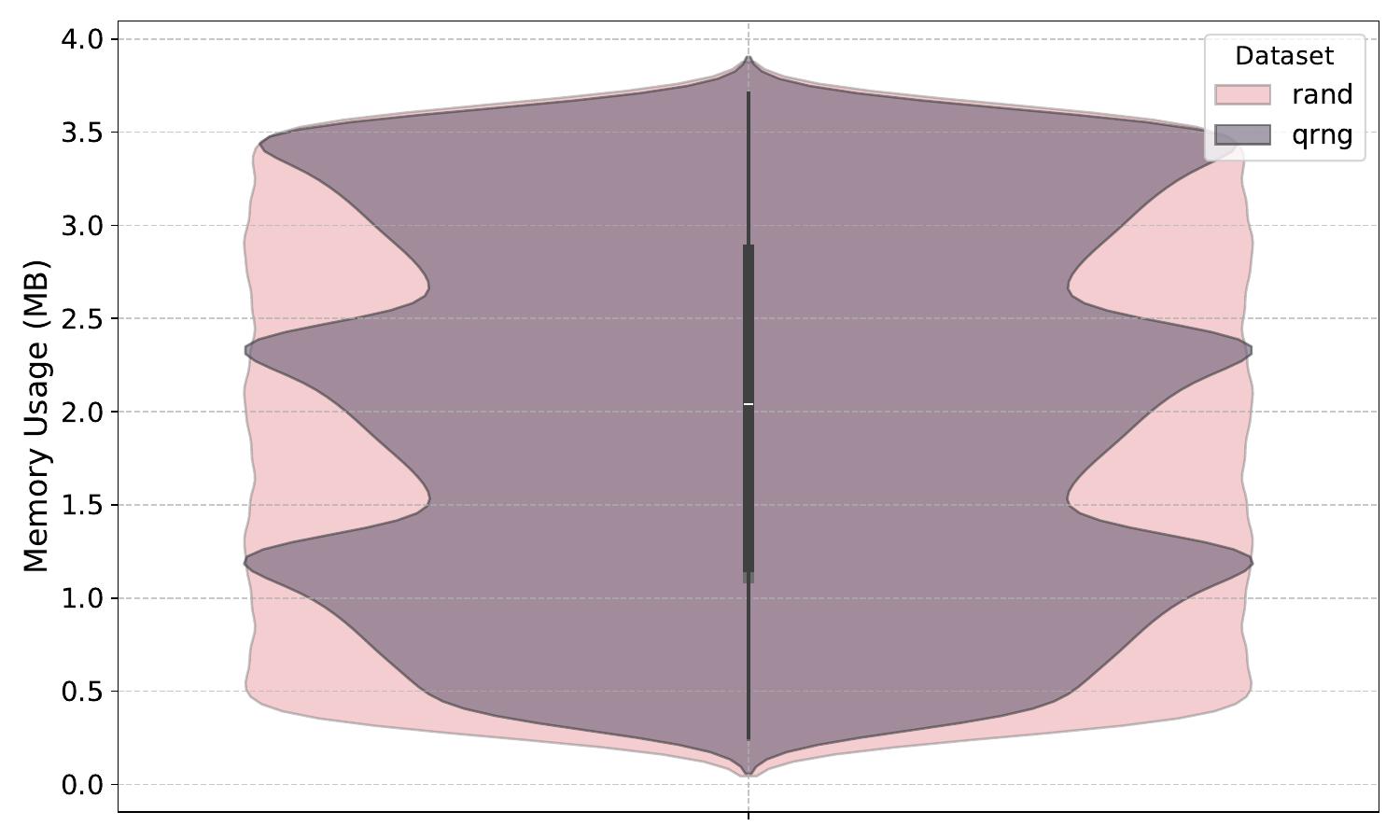}
    \caption{Memory Utilization}
    \label{fig:violin_mem}
\end{figure}

\textbf{Memory Utilization:}
The memory utilization of the qrng-based system remained relatively stable across most dataset sizes, averaging around 1.8 to 2.0 MB up to 8 million elements. However, a sharp drop to approximately 0.5 MB was observed at 10 million elements, indicating a potential optimization or anomaly in larger-scale memory management (Figure 5). In contrast, the Go-based VRF maintained consistent memory usage of around 1.0 to 1.2 MB across the entire dataset range, reflecting a more predictable memory profile suitable for applications where stability was critical (Figure 5).

\begin{figure}[htbp]
    \centering
    \includegraphics[width=1\linewidth]{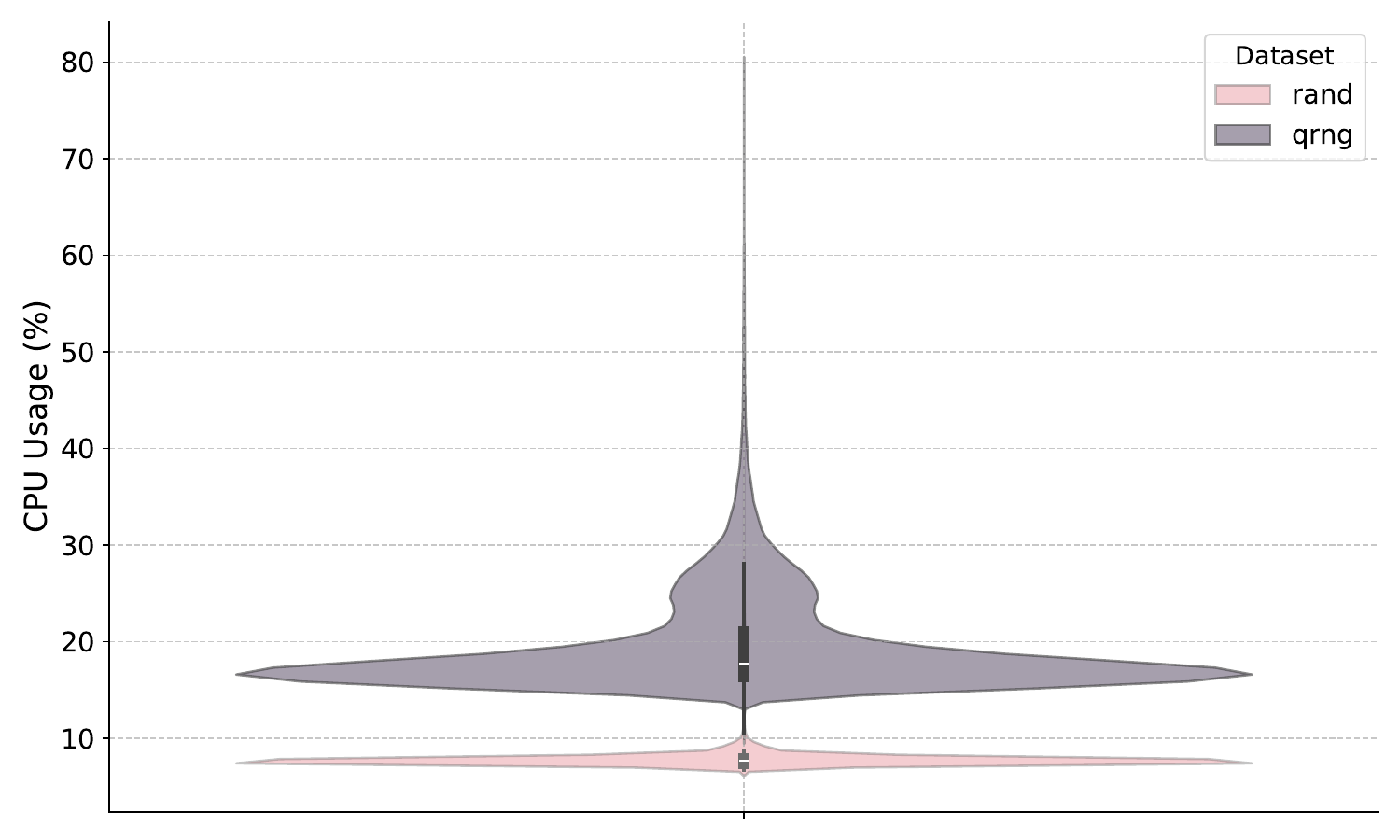}
    \caption{CPU Utilization}
    \label{fig:violin_cpu}
\end{figure}

\textbf{CPU Utilization:}  
The qrng system demonstrated higher CPU utilization than rand, ranging from 17\% to 30\%, with significant variability as the dataset size increased. This fluctuation suggested that qrng would require more processing power and exhibited less predictable CPU demand. Conversely, the rand maintained a much lower and consistent CPU usage, typically below 10\% (Figure 6), across all dataset sizes, highlighting its efficiency and suitability for environments with limited processing resources.

\subsection{Optimization Strategies for QRNG-Enhanced VRF Systems}

This section outlines a unified optimization strategy designed to mitigate the performance limitations in the QRNG-integrated VRF system. By focusing on both reducing I/O bottlenecks and enhancing concurrency, the proposed solution leverages advanced data handling techniques and parallel processing to optimize randomness generation and cryptographic operations.
The reliance on external QRNG-based randomness introduces significant I/O overhead, especially as the system scales and handles large cryptographic workloads. The current approach, which synchronously fetches QRNG data via the \texttt{BitReader} struct, incurs delays due to frequent file system access. This synchronous model, as seen in the \texttt{ReadBytes()} function, stalls cryptographic operations while waiting for the required randomness, resulting in latency spikes in high-throughput environments.
To resolve these issues, a combined strategy utilizing asynchronous I/O and multithreading is proposed. First, an asynchronous I/O model can be implemented by employing non-blocking file access techniques, allowing cryptographic processes to continue independently of the QRNG data retrieval. Go’s concurrency primitives, particularly \texttt{goroutines} \cite{strecansky2020hands}, can facilitate parallelism by enabling random data to be fetched in the background while elliptic curve computations proceed in parallel. This decouples QRNG randomness fetching from the cryptographic operations, significantly improving overall throughput.

Additionally, to minimize the frequency of file accesses, randomness retrieval should be optimized using a synchronous temporal merge (STM) strategy \cite{10.1145/2619092}. Instead of fetching small chunks of randomness for each cryptographic operation, larger blocks of QRNG data should be read and cached for multiple operations. STM, when applied to random read/write workloads, has demonstrated a 44\% improvement in mixed random workloads and a 1.25x to 1.44x increase in throughput under high concurrency (32 threads) compared to single-threaded approaches. This allows the system to aggregate multiple QRNG requests and serve them efficiently, minimizing file I/O operations and reducing the overall system latency caused by frequent disk reads.
Simultaneously, elliptic curve scalar multiplication—one of the most computationally intensive operations in the VRF system—should be parallelized. By leveraging Go’s multithreading capabilities, both scalar multiplication and randomness retrieval can be processed concurrently. In environments handling large-scale cryptographic tasks such as key generation (\texttt{GenKeyPair}) and proof generation (\texttt{GenerateVrf}), this parallelization ensures that CPU resources are fully utilized, thereby reducing system latency and increasing overall throughput.

\section{Conclusion}\label{sec:5}

This paper has proposed an innovative approach to verifiable randomness by integrating elliptic curve cryptography with quantum random number generators (QRNGs). The use of QRNGs significantly enhances security by providing true, non-deterministic randomness, making this system particularly suited for high-security applications, such as those requiring resistance to advanced cryptographic attacks. However, the integration of QRNGs also introduces challenges, particularly in terms of scalability and resource consumption in high-throughput environments, where continuous cryptographic proof generation is required.
Future research should delve deeper into post-quantum cryptography (PQC) to further enhance the security framework, focusing on the development of quantum-resistant algorithms. Additionally, exploring alternative QRNG techniques, optimizing system integration, and improving the overall efficiency and scalability of QRNG-based cryptographic systems will be essential steps toward making these technologies viable for mainstream use in a variety of secure, large-scale applications.

\section*{Acknowledgment}
\begin{itemize}
  
\item{\textbf{6G-life:}} The authors acknowledge the financial support by the Federal Ministry of Education and Research of Germany in the programme of “Souverän. Digital. Vernetzt.”. Joint project 6G-life, project identification number: 16KISK001K.
\item{\textbf{CeTI:}} Funded by the German Research Foundation (DFG, Deutsche Forschungsgemeinschaft) as part of Germany’s Excellence Strategy – EXC 2050/1 – Project ID 390696704 – Cluster of Excellence “Centre for Tactile Internet with Human-in-the-Loop” (CeTI) of Technische Universität Dresden. 
\item{\textbf{QD-CamNetz:}} The authors acknowledge the financial support by the Federal Ministry of Education and Research of Germany in the project QD-CamNetz, project identification number: 16KISQ076K.
\end{itemize}





\bibliographystyle{IEEEtran}
\bibliography{IEEEabrv,Bibliography}

\vfill


\end{document}